\documentstyle[prb,aps]{revtex}

\begin{document}
\title{Tuning gaps and phases of a two-subband system in a quantizing 
 magnetic field}
\author{ V.M. Apalkov and M.E. Portnoi\cite{byline}}
\address{School of Physics, University of Exeter, Stocker Road, 
 Exeter EX4 4QL, United Kingdom}

\maketitle

\begin{abstract}
 In this work we study the properties of a two-subband quasi-two-dimensional 
 electron system in a strong magnetic field when the electron filling 
 factor is equal to four. When the cyclotron energy is close to 
 the intersubband splitting the system can be mapped onto 
 a four-level electron system with an effective filling factor of two. 
 The ground state is either a ferromagnetic state or a spin-singlet state, 
 depending on the values of the inter-level splitting and Zeeman energy. 
  The boundaries between these phases are strongly influenced by the 
  inter-electron interaction. A significant exchange-mediated enhancement of 
 the excitation gap results in the suppression of the electron-phonon 
 interaction. The rate of absorption of non-equilibrium phonons is 
 calculated as a function 
 of Zeeman energy and inter-subband splitting. The phonon absorption rate 
 has two peaks as a function of intersubband splitting and has a step-like 
 structure as a function of Zeeman energy.

\end{abstract}


 During the last two decades the behavior of a two-dimensional (2D) electron 
 system in a quantizing magnetic field attracted a lot of theoretical and
 experimental interest. The peculiar properties of this system result from 
 its effective zero dimensionality, because the magnetic field normal
 to the 2D layer quenches the electron kinetic energy to a constant. As a 
 result the Hamiltonian of the system consists of the interaction part 
 only, and the system undergoes the transition into the incompressible 
 state at certain magnetic field values.
 The formation of the incompressible states results in the Fractional 
 Quantum Hall 
 Effect - one of the most interesting phenomena discovered in the 
 strongly correlated electron systems.\cite{QHE1,QHE2,QHE3} 
The only typical energy of the system is the characteristic 
 electron-electron Coulomb interaction energy, which is of the order of 
 $\varepsilon _C = e^2 /\kappa l $, 
 where $l$ is the magnetic length and $\kappa $ is the dielectric 
 constant. Introduction of a new degree of freedom into this system 
 results into new interesting phenomena when the characteristic energy of
 the new degree of freedom becomes smaller than the 
 Coulomb energy, $\varepsilon _C$. The electron spin is one of the 
 possible degrees of freedom. If the Zeeman energy is small enough then
 the transition from spin-polarized to spin-unpolarized ground state 
 occurs at electron filling factor $\nu _0 = 2/m$, where $m$ is 
 odd.\cite{spin} At filling factor $\nu _0 = 1/m$ a new type of 
 charged topological excitations, skyrmions, appears. 

 One can also introduce the degree of
 freedom in the direction normal to the 2D layer. This can be 
 realized in a double-layer system. In this case the phase diagram is
 driven by the interplay between the electron-electron Coulomb energy, 
 the tunnelling energy  between two layers and the Zeeman energy. 
 It was found that 
 at $\nu _0= 2$ the double-layer system can be found in 
 three different phases: fully spin-polarized ferromagnetic state, 
 spin-singlet state and canted antiferromagnetic state.\cite{dassarma}  

  In this paper we consider the system which is similar to a
 double-layer system but has a different geometry, which provides more 
 experimental possibilities to change the parameters of the system. 
 Namely, we study a single heterojunction in which the second degree 
 of freedom is introduced by the size quantization in the direction 
 perpendicular to the 2D layer. If the electron filling factor is greater 
 than two and the energy  $\Delta = \Delta_{12} -\hbar 
 \omega _c$ is smaller than the Coulomb energy, then the second subband 
 has to be taken into  account. Here $\hbar \omega _c$ 
 is the cyclotron energy and $\Delta_{12}$ is the inter-subband splitting. 
 In this case the second Landau level (Landau level number $n=1$) of 
 the first subband is close in energy to the first Landau level 
 ($n=0$) of the second subband. 
 Since the intersubband splitting $\Delta _{12}$ 
 has a weak dependence on the magnetic field, the separation 
 $\Delta  $ can  be changed by changing magnetic field $B$. 
 The influence of the second 
 subband on the optical properties of the magnetically-quantized 
 quasi-2D systems was observed in the different types of 
 experiments, e.g., in extrinsic radiative recombination 
 magnetospectroscopy\cite{plaut} and in the optically-detected cyclotron 
 resonance in tilted magnetic field.\cite{dw_e} 
 It was  proposed in Ref.~\onlinecite{misha1} that the 
 intersection between the levels can be also observed in the magnetic 
 field dependence of the phonon-mediated conductivity of the system. 
 At filling factor slightly greater than two, 
 the dissipative conductivity should reveal a double-peak structure as a 
 function of magnetic filed. A similar double-peak structure should be 
 observed in the other phonon spectroscopy 
 experiments.\cite{phonon1,phonon2,phonon3} 
 With increasing electron density the repulsion between the levels due 
 to electron-electron exchange interaction opens a gap in the inter-level 
 excitation spectra.  As a result the double-peak structure transforms 
 into a single-peak one.\cite{nu1} 

\begin{figure}
\begin{center}
\begin{picture}(110,70)
\put(0,0){\includegraphics{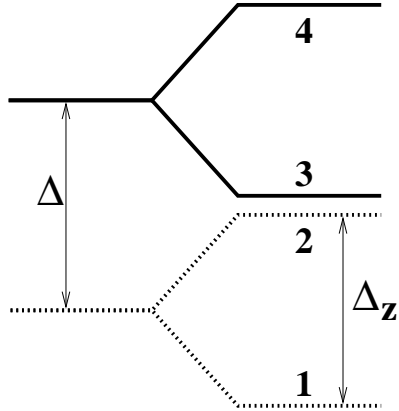}}
\end{picture}
\vspace*{4.0cm}
\caption{The single-electron energy levels of the two-subband system.
  The levels '1' and '2' belong to  the second Landau level ($n=1$)
  of the first subband with spins $S_z=1/2$ and $S_z=-1/2$, respectively. 
  The levels '3' and '4' belong to the first Landau level ($n=0$) of the 
 second subband with spins $S_z=1/2$ and $S_z=-1/2$, respectively. 
 The splitting $\Delta $ is the interlevel splitting, 
 $\Delta _z$ is the Zeeman energy.
}
\label{fig1}
\end{center}
\end{figure}

  In what follows we study the properties of the two-subband 
 system when the electron filling factor is equal to four, $\nu _0 =4$. 
 We assume that the Coulomb energy is much smaller than the cyclotron 
 energy. In this case the completely occupied 
 lowest Landau level of the first subband can be considered as a
 non-dynamical background.
 We will be interested only in a subsystem consisting of the second Landau 
 level ($n=1$) of the first subband and the first Landau level ($n=0$) of 
 the second subband, the spin degeneracy of both levels is lifted by the 
 Zeeman splitting $\Delta _z$. This four-level subsystem is schematically 
 shown in  Fig.~\ref{fig1}. For $\nu _0=4$ the effective 
 filling factor $\nu $ of this subsystem is equal to two, $\nu =2$.
 Under this mapping the system becomes similar to a
 double-layer system with the total filling factor $\nu =2$ and for its 
 analysis we follow the method of Ref.~\onlinecite{dassarma}. 
 
  For non-interacting electrons we can distinguish three cases: 
 1) if $\Delta > \Delta _z$ the states '1' and '2' are occupied, and  
 the ground state is a spin singlet state; 
 2) if $ -\Delta _z < \Delta < \Delta _z$ the states '1' and '3' are 
 occupied, and the ground state is a ferromagnetic state; 
 3) if $\Delta < - \Delta_z$ the states '3' and '4' are  occupied, and 
 the ground state is again a spin singlet state. 
 The boundaries between these phases are given by the equations: 
 $\Delta = \Delta_z$ and $\Delta = -\Delta _z$. 
 When the electron-electron interaction is taken 
 into account the transitions between spin singlet and ferromagnetic states 
 can occur through a new phase, for example through a canted antiferromagnetic 
 phase.\cite{dassarma} Our calculations show that in 
 the present system there is no intermediate phase and the 
 transitions from the singlet to the ferromagnetic state are sharp, 
 like in the non-interacting case.  

 In what follows we use the Coulomb energy $\varepsilon _C$ 
 as the unit of energy and magnetic length $l$ as the unit of length. 
 In the Landau gauge with the vector potential $\vec{A} = (0,Bx,0)$ the 
 eigenstates 
 of a single-electron Hamiltonian are characterized by the Landau level
 number, $n$; $y$-component of the momentum, $k_y$; 
 the electron subband number,
 $\mu =1$ or $\mu =2$, and the $z$-projection of the electron spin, 
 $\sigma = 2S_z=\pm 1$:
\begin{equation}
\psi _{n,k_y, \mu, \sigma} (x,y,z) = \chi _{\mu} (z) \xi _{\sigma }
 \frac{e^{i k_y y}}
{\sqrt{L_y}} \phi _n (x-k_y )  \mbox{\hspace{3mm},}
\label{eq1}
\end{equation}
where $\chi _{\mu } (z)$ is the envelope wavefunction of the $\mu $th subband;
 $\xi _{\sigma }$ is the spin part of the wavefunction; $\phi _n (x)$ is the 
 $n$th harmonic oscillator function. 

   The Hamiltonian of the interacting electron system is
\begin{eqnarray}
H & & =  \sum_{n \mu \sigma k_y} \hbar \omega _c \left(n+\frac{1}{2} \right)
   C^{+} _{n k_y \mu \sigma}
 C _{n k_y \mu \sigma}+ \frac{\Delta }{2} \sum_{n \mu \sigma k_y} (2\mu -3) 
   C^{+} _{n k_y \mu \sigma}
 C _{n k_y \mu \sigma} +                    \nonumber  \\
& & + \frac{\Delta _z}{2} \sum_{n \mu \sigma k_y} \sigma
 C^{+} _{n k_y \mu \sigma} C _{n k_y \mu \sigma}      
 + \frac{1}{2} \sum _{ \left\{n \right\} } \sum_{\sigma _1 \sigma _2} 
                   \sum_{ \left\{\mu \right\} } 
  \sum_{q_x, q_y} \tilde{V}^{(n_1 n_4 n_3 n_2)}_{(\mu _1 \mu _4 
                                              \mu _3 \mu _2)} 
 (\hat{q}) 
  \times   \nonumber \\  
 & & \times  \sum _{k_1,k_2} e^{i q_x (k_1-k_2)} C^{+} _{n_1,k_1+q_y, \mu_1
 \sigma_1}C^{+} _{n_2, k_2, \mu_2 \sigma_2}  C _{n_3, k_2+q_y, \mu_3 \sigma_2} 
C^{+} _{n_4, k_1, \mu_4 \sigma_1}  \mbox{\hspace{3mm},}
\label{eq2}
\end{eqnarray}
  where $ C^{+} _{n, k, \mu \sigma}$ and $ C _{n, k, \mu \sigma } $ are the 
 creation and annihilation operators of the electron in the state 
 $\psi _{n,k_y, \mu, \sigma}$. In Eq.~(\ref{eq2}) we use the notations:
\begin{equation}
\tilde{V}^{(n_1 n_4 n_3 n_2)}_{(i_1 i_4 i_3 i_2)}(\hat{q}) = \frac{1}{q}
         F_{i_1 i_4 i_3 i_2}(q)  G _{n_1 n_4}(\hat{q}) 
       G _{n_3 n_2}(\hat{q}^*)   
      \exp\left ( -\frac{q^2}{2} \right)  \mbox{\hspace{3mm},}
\label{eq3}
\end{equation}
where\cite{kallin}
\[
G_{n_1 n_2} (\hat{q}) = \left( \frac{n_1!}{n_2 !} \right)^{1/2} 
              \left( \frac{-i \hat{q}}{\sqrt{2} } \right)^{n_2 - n_1}
           L_{n_1}^{n_2-n_1} \left( \frac{ q^2}{2} \right) \mbox{\hspace{5mm},}
\]
$\hat{q}=q_x + iq_y$, $q = |\hat{q}|$ and $L^{m}_{n}$ is a generalized 
Laguerre polynomial. The modification 
of the Coulomb interaction due to the finite extent of the electron 
 wavefunctions in $z$-direction is given by:
\[
F_{i_1 i_4 i_2 i_3} (q) = \int _{0}^{\infty } \int _{0}^{\infty }
   dz_1 dz_2 e^{-q|z_1-z_2|}\chi _{i_1} (z_1) \chi _{i_4} (z_1)
                                                 \chi _{i_2} (z_2) 
     \chi _{i_3} (z_2)
          \mbox{\hspace{3mm}.}          
\]
 We use the Fang-Howard approximation\cite{ando} for the envelope 
 wavefunctions of the electrons in the first and the second subbands: 
\begin{eqnarray*}
& & \chi _1(z) = \sqrt{\frac{b^3}{2}} z \exp\left( -\frac{1}{2} bz \right)  \\
& & \chi _2(z) = \sqrt{\frac{b^5}{6}} z\left( z- \frac{3}{b} \right) 
                          \exp\left( -\frac{1}{2} bz \right) 
  \mbox{\hspace{3mm}.}
\end{eqnarray*}

 We follow the standard Hartree-Fock method, assuming a non-zero average of 
$\left< C^{+} _{k_1, \mu _1 \sigma _1} C _{k_2, \mu _2 \sigma _2}\right>$
 over the ground state. The 
 specific feature of the given problem, which differs it from the 
 double-layer system, is that the non-zero paring in (\ref{eq2}) can 
 occur between the states with the different value of $k_y$, because 
 in our case the two-level system is formed by the different Landau 
  levels.\cite{quinn}
 For each value of $k_y = k$ we introduce the new wavefunctions, which are 
 the eigenfunctions  of the Hartree-Fock Hamiltonian. The  creation and 
 annihilation operators corresponding to the new wavefunctions 
 are $a^{+} _{k,i }$, $a  _{k,i }$, where $i=1,2,3$ or $4$. 
These functions
 are related to the original ones by the matrix $\gamma _{i,y}$: 
\begin{equation}
 \begin{array}{l}
C_{k,1,-1} = \sum_{i=1}^{4} \gamma _{1i}a _{k,i} \\
C_{k,1,1} = \sum_{i=1}^{4} \gamma _{2i}a _{k,i} \\
C_{k+Q,2,-1} = \sum_{i=1}^{4} \gamma _{3i}a _{k,i} \\ 
C_{k+Q,2,1} = \sum_{i=1}^{4} \gamma _{4i}a _{k,i}      \mbox{\hspace{3mm}.} 
\end{array}
\label{eq4}
\end{equation}
The average of the introduced functions over the ground state is 
 $\left<  a^{+}_{k,i}a_{k,j} \right> = 
 \delta _{ij}(\delta _{i1}+ \delta _{i2}) $. This means that only the 
 states  with the lowest  energies ($i=1$ and $i=2$) are occupied. 
Substituting  Eq.(\ref{eq4}) into the Hamiltonian (\ref{eq2}) 
 we obtain the Hartree-Fock Hamiltonian in the form 
 of a four-by-four matrix. In the  basis 
$( C_{k,1,-1}, C_{k,1,1}, C_{k+Q,2,-1}, C_{k+Q,2,1})$ the elements of 
 this matrix are
\begin{eqnarray*}
& & H^{HF}_{11} = -\frac{1}{2}(\Delta +\Delta_z)-\beta _{11}\epsilon ^{11}_{11}
                         -\beta  _{33} \epsilon^{01}_{12}   \\
& & H^{HF}_{22} = -\frac{1}{2}(\Delta -\Delta_z)-\beta _{22}\epsilon ^{11}_{11}
                         -\beta  _{44} \epsilon^{01}_{12}    \\ 
& & H^{HF}_{33} = \frac{1}{2}(\Delta  -\Delta_z)-\beta _{33} 
                                                          \epsilon^{00}_{22} 
                       - \beta  _{11} \epsilon^{01}_{12}  \\
& & H^{HF}_{44} = \frac{1}{2}(\Delta +\Delta_z)-\beta _{44} \epsilon^{00}{22}  
                        -\beta  _{44} \epsilon^{01}_{12}  \\
& & H^{HF}_{12} = -\beta _{12} \epsilon ^{11}_{11} - (\beta _{23}-\beta _{14}) 
                     V^{(0100)}_{(2111)}(Q) - \beta _{34} \epsilon^{01}_{12} \\
& & H^{HF}_{13} = -\beta _{13} \left( 
                                 V^{(0011)}_{(1122)}(0)+V^{(0101)}_{(1212)}(Q) 
\right) \\ 
& & H^{HF}_{14} = -\beta _{23}V^{(0101)}_{(1212)}(Q) - \beta _{14} 
V^{(0011)}_{(1122)}(0)   \\
& & H^{HF}_{23} =-\beta _{14}V^{(0101)}_{(1212)}(Q) - \beta _{23}  
V^{(0011)}_{(1122)}(0) \\ 
& & H^{HF}_{24} =  -\beta _{24}\left( 
                                V^{(0011)}_{(1122)}(0)+V^{(0101)}_{(1212)}(Q) 
\right)  \\
& & H^{HF}_{34} = -\beta _{34}\epsilon^{00}_{22} -(\beta  _{23}-\beta _{14})
                   V^{(0111)}_{(1222)}(Q) - \beta _{12} \epsilon^{01}_{12}   
                                              \mbox{\hspace{3mm},}
\end{eqnarray*}
where $\beta _{ij} = \gamma _{i1}\gamma _{j1} + \gamma _{i2}\gamma _{j2} $,
\begin{equation}
V^{(i_1 i_2 i_3 i_4)}_{(j_1 j_2 j_3 j_4)}(q) = 
    \int _0 ^{\infty }dk  F_{j_1 j_2 j_3 j_4}(k) \left| G _{i_1 i_2}(k) 
\right| 
      \left| G _{i_3 i_4}(k) \right| J_{\left| i_1 +i_3 -i_2-i_4 
\right| }(kq)  
      \exp\left ( -\frac{k^2}{2} \right)          \mbox{\hspace{3mm},}
\label{eq5}
\end{equation}
\[
\epsilon ^{i_1 i_2}_{j_1 j_2} = V^{(i_1 i_2 i_1 i_2)}_{(j_1 j_2 j_1 j_2)}(0)
                                  \mbox{\hspace{3mm},}
\]
 $J_m$ is the Bessel function of the $m$th order, and 
$\epsilon ^{i_1 i_2}_{j_1 j_2}$ is the exchange energy of an electron 
 in the $i_1$th Landau level of the $j_1$th subband interacting with  
 electrons of the same spin in the filled $i_2$th Landau level of the 
 $j_2$th subband. 

\begin{figure}
\begin{center}
\begin{picture}(110,70)
\put(0,0){\includegraphics{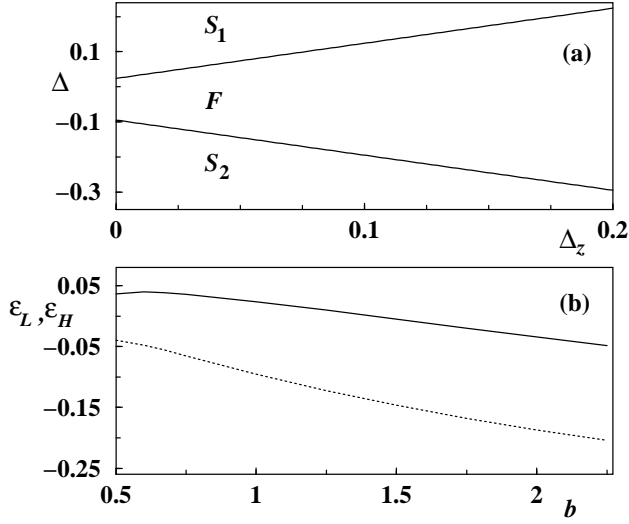}}
\end{picture}
\vspace*{5.0cm}
\caption{(a) The phase diagram of the four-level system 
 for $b=1$. The phases $S_1$ and $S_2$ are spin singlet phases, 
 the phase $F$ is a ferromagnetic phase.  
   (b) The parameters $\epsilon _H$ and $\epsilon _L$
 (Eqs.(6)-(7)) of the phase diagram are plotted as the functions of
 $~b$. All energies are in units of $\varepsilon _C$, $~b$ 
 is in units of $1/l$.  
}
\label{fig2}
\end{center}
\end{figure}

Taking into account that the eigenvectors of the Hartree-Fock 
 matrix $a_{k,i}$ are related to $C_{k,\mu , \sigma}$ 
 by the matrix $\gamma _{ij}$  
 (Eq.~(\ref{eq4})), we obtain the self-consistent system of equations for 
 $\gamma _{ij}$.  The parameter $Q$ can be found by minimizing the 
 Hartree-Fock energy. We have found the solution of the
  system of equations for $\gamma _{ij}$ numerically. For all values of 
  $\Delta $ and $\Delta _z$ the ground state of the system is one of 
 the non-interacting phases (spin singlet or ferromagnetic), described above.
 The transitions between the phases are sharp. 
 The interaction  modifies the phase boundaries only:
\begin{eqnarray}
& & \Delta  = \Delta _z + \frac{1}{2} \left( \epsilon _{22}^{00}- 
             \epsilon _{11}^{11} \right) + \epsilon _{12}^{00}
             - \epsilon _{11}^{01} + \epsilon _{12}^{01} =
 \Delta _z + \epsilon _{H} = \Delta _H             \mbox{\hspace{3mm},}      
                                                   \label{eq6}  \\
& &  \Delta = -\Delta _z + \frac{1}{2} \left( \epsilon _{22}^{00}- 
             \epsilon _{11}^{11} \right) + \epsilon _{12}^{00}
             - \epsilon _{11}^{01} - \epsilon _{12}^{01}
   = -\Delta _z +\epsilon _{L} = \Delta _L   \mbox{\hspace{3mm},}
                                                   \label{eq7}
\end{eqnarray}
 where $\Delta _H $ and $\Delta _L$ denote the higher and lower 
 values of splitting $\Delta $ at the boundary of
 the ferromagnetic phase, see Fig.~\ref{fig2}(a). 
In Fig.~\ref{fig2}(a) the phase diagram is shown for $b=1$. The phases $S_1$ 
 and $S_2$ are the  spin-singlet phases and the phase $F$ is the 
 ferromagnetic phase. The single-electron states '1' \& '2', '1' \& 
 '3', and '3' \& '4' (Fig.~\ref{fig1}) are occupied in 
the phases $S_1$, $F$, and $S_2$, 
 respectively. The electron-electron interaction results in the
 splitting of the boundaries between different phases, even when 
 the Zeeman energy is equal to zero. 
 For $\Delta _z=0$ the splitting is equal to $ \epsilon _{H} - 
 \epsilon _{L} = 2 \epsilon _{12}^{01}$, where $ \epsilon _{H}$ and 
 $ \epsilon _{L}$ are introduced in Eqs.~(\ref{eq6}) and (\ref{eq7}).
 In Fig.~\ref{fig2}(b) the values of the lower ($ \epsilon _{L}$) and upper 
 ($\epsilon _{H}$) phase boundaries at 
 $\Delta _z =0$ are shown as the function of the Fang-Howard parameter $b$, 
 where $1/b$ is proportional to the characteristic width of the 
 heterojunction in units of magnetic length. Both  values $\epsilon _{H}$ and  
 $ \epsilon _{L}$ decrease with increasing $b$. 

 We are mostly interested in the case when the intersubband
 splitting is close to the cyclotron energy. This gives the 
 restriction of the values of the parameter $b$. This parameter 
 should be close to the inverse magnetic length. However,
 the exact relation between $b$ and $l$ for $\Delta =0$ depends on the 
 actual value of $b$. Therefore we found it appropriate to calculate 
 the $b$-dependencies of the main parameters of the system.

 The two-subband system can be studied experimentally by changing magnetic
 field when keeping the fixed value of the filling factor. If the 
 magnetic field $\vec{B}$ is normal to the heterojunction, 
 then both $-\Delta $ and $\Delta _z$ increase
 linearly with increasing $B$. In this case only the transition from 
 the phase $S_1$ to the phase $F$ can be observed. 
 Another possibility is to make the experiments in the tilted magnetic 
 field with the fixed value of its normal component. In this 
 case $\Delta $ is a constant, because it is proportional to 
 the normal component of the magnetic field, but $\Delta _z$ 
 changes linearly with the total magnetic field. Depending 
 on the initial value of $\Delta $, one should observe the
 transition from $S_1$ to $F$ or from $S_2$ to $F$ phases. 
 However, there is a finite region of $\Delta $ in which
 no transitions can be found and the system is always in the 
 phase $F$. This region has a non-zero range even for zero $\Delta _z$ when 
 its width is $2 \epsilon _{12}^{01}$.

 Now we shall discuss the possibility to detect the different phases by 
 the acoustic-phonon spectroscopy.\cite{phonon1,phonon2,phonon3}
 We consider the absorption of non-equilibrium phonons by the discussed  
 quasi-2D system. The electron-phonon interaction is proportional to the 
 electron density operator.\cite{levinson} 
 During the act of the phonon absorption the electron system undergoes 
 the transition from the ground state to the excited state, corresponding 
 to the density fluctuations. When the system is in the
 ferromagnetic phase ($F$), there are spin reversal excitations only. 
 In this case the phonon absorption is forbidden.  
 The collective excitation spectra in the singlet phases $S_1$ and $S_2$ 
 can be found from the poles of the density-density  correlation 
 function\cite{kallin}, which are given by the expressions:
\begin{equation}
E(q) = \Delta + 2\tilde{V} _{1212}^{0101}(q) - V_{1122}^{0011}(q) -
     \epsilon _{12}^{01} - \epsilon _{12}^{00} + \epsilon _{11}^{11}
     + \epsilon _{11}^{01}
\label{eq8}
\end{equation}
for the phase $S_1$ and
\begin{equation}
E(q) = -\Delta + 2\tilde{V} _{1212}^{0101}(q) - V_{1122}^{0011}(q) -
     \epsilon _{12}^{01} - \epsilon _{11}^{01} + \epsilon _{22}^{00}
     + \epsilon _{12}^{00}
\label{eq9}
\end{equation}
for the phase $S_2$, where Eqs.~(\ref{eq3}) and (\ref{eq5}) were used. 
One can rewrite Eqs.~(\ref{eq8}) and (\ref{eq9}) in the form,
which can describe both cases simultaneously:
\begin{equation}
E(q) = |\delta \Delta| +\Delta _z + 
  2\tilde{V} _{1212}^{0101}(q) - V_{1122}^{0011}(q) + \frac{1}{2} \left(
     \epsilon _{22}^{00} + \epsilon _{11}^{11} \right)
              \mbox{\hspace{3mm},}
\label{eq10}
\end{equation}
where $\delta \Delta = \Delta -\Delta _H$ for the phase $S_1$ and
 $\delta \Delta = - \Delta + \Delta _L$ for the phase $S_2$.  

\begin{figure}
\begin{center}
\begin{picture}(110,70)
\put(0,0){\includegraphics{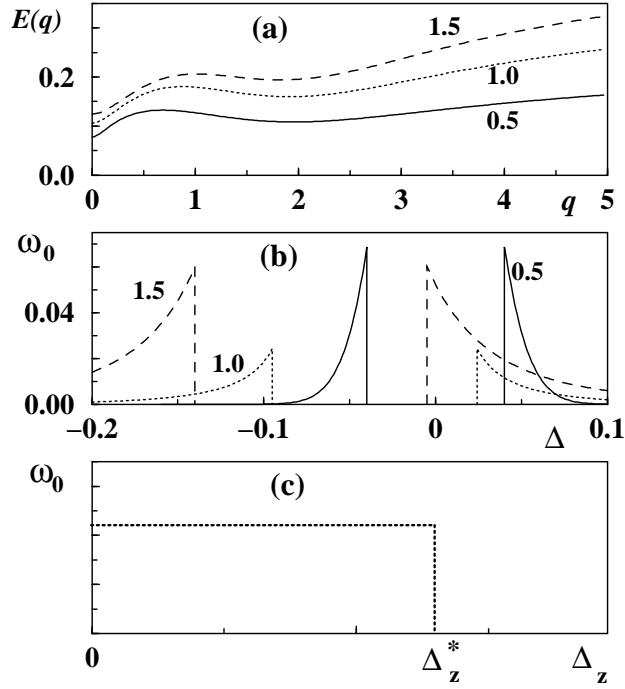}}
\end{picture}
\vspace*{8.0cm}
\caption{ (a) The excitation dispersion  $E(q)$ (Eq. (10)) for 
 $\delta \Delta = 0= \Delta _z$. The numbers near
 the  lines show the values of the parameter $b$, where 
 $b$ and $q$ are in units of $1/l$, $E(q)$ is in units 
 of $\varepsilon _C$.
   (b) The phonon absorption rate $\omega _0$ as a function 
 of $\Delta $ for $\Delta _z = 0$. Solid, dotted and
 dashed lines are for $b=0.5$, $1.0$ and $1.5$, 
 respectively. The data for  $b=1.0$ and $1.5$ are
 multiplied by 10. The absorption rate $\omega_0$ is in 
 units of $10^{10}~$s$^{-1}$, $\Delta $ is in units of 
 $\varepsilon _C$.
   (c) Phonon absorption rate $\omega _0$ as a function of
 $\Delta _z$ for constant $\Delta $. 
 Here $\omega_0$ and $\Delta _z$ are in arbitrary
 units. The critical Zeeman energy $\Delta _z^*$ corresponds to 
 the $S_1$$ \rightarrow $$F$  or $S_2$$\rightarrow $$F$ 
 phase transitions.
}
\label{fig3}
\end{center}
\end{figure}

The energy spectra $E(q)$ are shown in Fig.~\ref{fig3}(a) 
for $\delta \Delta=0$ and $\Delta _z=0$, for several different values of 
parameter $b$. One can see that  there is a finite gap for all 
 values of momentum $q$. This means that the gap exists for any 
 value of $\Delta $ and $\Delta _z$.
 With increasing the spreading of the electron wavefunction in $z$ direction 
 (decreasing $b$) the gap becomes smaller, which results from the 
  decreasing of the effective 2D electron-electron interaction. 
   The existence of the interaction-induced finite excitation 
 gap in the two-level system for any value of the inter-level splitting  
 was demonstrated for the system with  filling factor $\nu =1$ in 
 Ref.~\onlinecite{nu1}. The effect of the strong renormalization of the 
 excitation energy due to the inter-electron interaction is also known
 for spin excitations as an interaction enhancement of the $g$-factor 
 in 2D systems.\cite{g1,g2,g3,g4} 

  The electron-phonon interaction Hamiltonian has the form:
\begin{equation}
H_{e-ph} = - \sum _{j, \vec{Q}} \frac{M_j (\vec{Q})}{\sqrt{V}} Z(q_z) \left[ 
          \rho(\vec{q}) \hat{d}^{+}_j (\vec{Q}) + 
          \rho(-\vec{q}) \hat{d} _j (\vec{Q}) \right] \mbox{\hspace{3mm},}
\label{eq11}
\end{equation}
where the isotropic Debye approximation is used with the linear 
 dependence  of the phonon frequency on the wave vector:
$\omega _{j} (K) = s_{j} K $,
$s_{j}$ being speed of sound; $j$ is labeling the phonon modes,
 $j=1$ for longitudinal mode and $j=2,3$ for two transverse modes.
In Eq.~(\ref{eq11}) we labeled the three dimensional (3D) 
 phonon wave vector by the capital letter, $\vec{Q}$, 
 and its projections by the 
 corresponding small letters, $\vec{Q}=(\vec{q},q_z)$.
 The creation operator of a phonon in the $j$th mode is  denoted 
 as  $\hat{d}^{+}_j$, $V$ is a normalization volume, 
 $\rho (\vec{q})$ is the two-dimensional electron density operator,
 $M_j (\vec{Q})$ is the matrix element of the electron-phonon
 interaction, which is determined by the deformation potential and 
 piezoelectric coupling:\cite{benedict}
\begin{equation}
M_j (\vec{Q}) = \sqrt{\frac {\hbar }{2 \rho_0 s Q}} 
  \left[ -\beta 
           \frac{Q_x Q_y \xi _{j,z} +Q_y Q_z \xi _{j,x} +
                           Q_z Q_x \xi _{j,y}}{Q^2}
    - i \Xi _0 (\vec{\xi}_{j} \cdot \vec{Q} ) \right]
  \mbox{\hspace{3mm},}
\label{eq12}
\end{equation}
 where $\rho _0 $ is the mass density, 
 $\beta $ and $\Xi _0$ are the parameters of piezoelectric
 and deformation potential couplings, $\vec{\xi}_j$ is the polarization
 vector of the $j$th phonon mode. The parameters of GaAs are used in our
 calculations.

 Because the phonon-assisted transitions are only allowed between the 
 different subbands, the form factor $Z(q_z)$ is
\begin{equation}
Z(q_z) = \int dz e^{i q_z z} \chi_1(z) \chi_2(z) \mbox{\hspace{3mm}.}
\label{eq13}
\end{equation}

 At low temperature, $k_B T\ll E(q)$, the rate of absorption of 
 non-equilibrium phonons can be found from the expression:\cite{nu1,benedict}
\begin{equation}
\omega _{abs} = \frac{2\pi}{\hbar}  \sum _{j}
\int \frac{d \vec{Q}}{(2\pi )^3} \delta (E(q) - s_j Q) n_{j}(\vec{Q})
              \left| M_j (\vec{Q}) Z(q_z) \right|^2 R_{01}(q)
     \mbox{\hspace{3mm},}
\label{eq14}
\end{equation}
 where $R_{01}(q) = (q^2/2) \exp(-q^2/2)$, $ n_{j}(\vec{Q})$ is the
 phonon distribution function.

  One can see how the appearance of the excitation gap suppresses 
 the phonon absorption. The phonon absorption process requires 
 the conservation of energy and in-plane momentum, 
 $s_j \sqrt{q^2+q_z^2}=E(q)$, where 
 the characteristic excitation gap, $E(q)$, is greater than 
 $0.1$ for $b\sim 1$ (see Fig.~\ref{fig3}(a)).
 The GaAs speed of sound in units of
 $l\epsilon _C /\hbar$ is about 0.03 for longitudinal and 0.02 
 for transverse phonons.
 Therefore, the large value of $\sqrt{q^2+q_z^2} $ is required 
 to satisfy the energy conservation law. However, 
 the factors $R_{01}(q)$ and $Z(q_z)$ in Eq.~(\ref{eq14}) make
 the phonon absorption to be small at $q> 1$ or $q_z > b$. Thus, 
 the phonon absorption becomes strongly suppressed comparing 
 to the non-interacting case, in which the excitation 
 energy $\Delta $ is allowed to have any value.
 For the non-interacting electrons the absorption rate has two
 maxima at $|\Delta |\sim \hbar s_j/l$.\cite{misha1}

 From Eq.~(\ref{eq14}) we calculate the normalized phonon 
 absorption rate, $\omega _0  = \omega _{abs}/n(E(q_0))$, 
 where $q_0$ is a characteristic momentum, here we use $q_0 = 1$.
 In Fig.~\ref{fig3}(b) the rate $\omega_0$ is shown as a function of 
 $\Delta $ for $\Delta _z=0$ and different values of $b$. The 
 absorption rate has a two-peak structure. The gap 
 between the peaks corresponds to the phase $F$ of the
 system, in which the phonons can not be absorbed. The 
 absorption  rates for 
 $b=1$ and $b=1.5$ are multiplied by 10. With increasing $b$ 
 the collective excitation gap increases
 (see Fig.~\ref{fig3}(a)), which results in a rapid 
 decrease of the absorption rate. One can see this tendency for 
 $b=0.5$ and $b=1$. However, with increasing $b$ 
 the phonons with the larger values of $q_z$ can be absorbed,
 which tends to increase the  absorption rate. The latter effect 
 becomes more pronounced for large $b$ when the excitation gap 
 has a weaker dependence on $b$ (see Fig.~\ref{fig3}(a)). 
 The competition between these two effects results in the 
 slight increase of the absorption rate for $b=1.5$. 

  In the tilted magnetic field experiments, when the Zeeman 
 energy $\Delta _z$ is changed while the inter-level 
 splitting $\Delta $ is kept fixed, the absorption rate 
 remains constant. 
 For the singlet phase $S_1$ or $S_2$ the absorption rate 
 is given by Eqs.~(\ref{eq8}), (\ref{eq9}), (\ref{eq14})
 and for the ferromagnetic  phase $F$ the absorption rate is zero. 
 The dependence of $\omega _0$ on $\Delta _z$ is shown 
 schematically in Fig.~\ref{fig3}(c), where the critical 
 value of Zeeman splitting, $\Delta _z^*$, corresponds to 
 $S_1$-$F$ or $S_2$-$F$ phase transition.   
    
 In conclusion, we considered the two-subband electron system 
 with the total filling factor four. The system was mapped onto the 
 four-level electron system with the effective filling factor two. 
 The separation between the levels belonging to the different 
 subbands is proportional to the cyclotron energy and can be 
 changed by changing the magnetic field. 
 The system exists in one of the three different phases: 
 one is ferromagnetic and two are spin-singlet phases. 
 The electron-electron interaction does not create new phases. 
 However, it renormalizes the phase 
 boundaries. The excitation spectrum has the gap for any value of 
 the inter-level splitting $\Delta $ and Zeeman splitting $\Delta _z$. 
 This results in the strong suppression of the electron-phonon interaction. 
 The rate of the phonon absorption by the considered quasi-2D 
 electron system has a double-peak structure as a function of 
 level splitting and a 
 step-like structure as a function of Zeeman splitting. 

 The authors  acknowledge the support by the UK EPSRC.

\end{document}